\documentclass[preprint,authoryear,12pt]{elsarticle}

\usepackage{amssymb}

\journal{Studies in Hist. Philos. Mod. Phys.}

\begin{document}

\begin{frontmatter}

\title{Non-standard Models and the Sociology of Cosmology\footnote{Paper accepted for publication
in {\it Studies in History and Philosophy of Modern Physics (SPHMP)} for the special issue on ``Philosophy of Cosmology'' (2014). Sentences in {\it italics} [apart from the titles of books or journals or titles
of subsections, which are always in italics] are published only in this {\it arXiv.org} version and
they have been removed or substituted in the version of the journal {\it SPHMP}, since the editors and referees of this paper asked me to do this in order to the article be accepted in the journal. Here, I keep some of the original sentences and paragraphs because I prefer this version rather than the filtered one.}}

\author{Mart\'\i n L\'opez-Corredoira}

\address{Instituto de Astrof\'\i sica de Canarias, E-38200, La 
Laguna, Tenerife, Spain \\ Departamento de Astrof\'\i sica, Univ. La Laguna,
E-38206 La Laguna, Tenerife, Spain}

\begin{abstract}
I review some theoretical ideas in cosmology different from the standard ``Big Bang'': 
the quasi-steady state model, the plasma cosmology model, non-cosmological redshifts, 
alternatives to non-baryonic dark matter and/or dark energy, and others. 
Cosmologists do not usually work within the framework of alternative cosmologies because they feel that
these are not at present as competitive as the standard model. Certainly, they are not so 
developed, and they are not so developed because cosmologists do not work on them. It is a 
vicious circle. The fact that most cosmologists do not pay them any attention and only 
dedicate their research time to the standard model is to a great extent due to a sociological 
phenomenon (the ``snowball effect'' or ``groupthink''). We might well wonder whether cosmology, our 
knowledge of the Universe as a whole, is a science like other fields of physics or a predominant ideology. 

\end{abstract}


\end{frontmatter}

\section{Introduction}

The present-day standard model of cosmology (the ``Big Bang'') gives us a representation of 
a cosmos whose dynamics is dominated by gravity (from general relativity), with a finite 
lifetime, large scales homogeneity, expansion and a hot initial state, together with 
other elements necessary to avoid certain inconsistencies with the observations 
(inflation, non-baryonic dark matter, dark energy, etc.). Although the Big Bang is the 
most commonly accepted theory, it is not the only possible representation of the Cosmos. 
In the last $\sim$90 years ---such is the brief history of the branch of science called 
cosmology--- there have been plenty of other proposals. I describe them in
\S \ref{.alt} of this paper.

Cosmologists do not usually work within the framework of alternative cosmologies because they feel
these are not at present as competitive as the standard model. Certainly, they are not so 
developed, and they are not so developed because cosmologists do not work on them. It is a 
vicious circle. The fact that most cosmologists do not pay them any attention and only 
dedicate their research time to the standard model is to a great extent due to a sociological 
phenomenon. In a second part of the paper, \S \ref{.soc} and \S \ref{.lim}, I will discuss 
the sociological aspects  related to cosmology and the debate on the different theories. 

\section{Alternative models}
\label{.alt}

Although the standard model (``Big Bang'') is the most well known and commonly accepted theory 
of cosmology, it is not the only possible representation of the Cosmos, and it is not clear 
that it is the right one, not even in an approximate way 
(for a discussion of some of its problems see L\'opez-Corredoira,
2003, and see also below in \S \ref{.caveats}). 
There were and there are many other alternative approaches to our understanding of the Universe 
as a whole. Among them, because of its historical importance and impact, the 
quasi-steady state model and plasma cosmology are significant examples. There are many other
examples too. I will give a brief description of them in this section. I will not give a complete 
list of models, but this sample is large enough to give an idea of what theoretical approaches 
are being discussed in cosmology from heterodox standpoints: either from dissidence 
with respect to the standard model, or dissidence with respect to the dominant
dissident theories.

\subsection{Quasi-Steady State Cosmology}

The theory (better call it a hypothesis) which is called nowadays the ``quasi-steady state 
cosmology'' (QSSC) was indeed first called the ``steady state theory''. Hoyle (1948), and independently 
Bondi \& Gold (1948), proposed the hypothesis of the steady state in which, contrary to the Big Bang 
approach, there was no beginning of the Universe. The Universe is expanding, it is
eternal and the homogeneous distribution of matter is being created at a rate of $10^{-24}$ 
baryon\-/cm$^3$\-/s, instead of the unique moment of creation in
the Big Bang. The perfect cosmological principle of a Universe which is observed to be the same from 
anywhere and at any time is followed in this model, whereas the standard model only gives a 
cosmological principle in space but not in time. There is no evolution. The Universe remains always 
the same. Newly created matter forms new galaxies which substitute those that are swept 
away by the expansion.

Fred Hoyle (1915--2001) inadvertently baptised the rival theory:
he dub\-bed  the primaeval atom theory of Gamow and coworkers\footnote{George 
Gamow (1904--1968) and one of his students, Ralph Alpher, published  a paper in 1948.
Gamow, who had certain sense of humour, decided to put the reputed physicist 
Hans Bethe as second author, even though he had not participated in the
development of the paper. Bethe was amused, so the result was a paper by
Alpher, Bethe and Gamow (to rhyme with ``alpha, beta and gamma''). Later, R. C. Herman joined the 
research team, but---according to Gamow---he refused stubbornly to change his name to ``Delter''.}  the ``Big Bang''
in order to ridicule it. However,  the name caught on.
During the '50s, both theories held their ground. While there were attempts 
to explain the abundances of the chemical elements with Gamow et al.'s theory,  the Steady State Theory
also provided plausible explanations.
E. M. Burbidge et al. (1957) explained the abundances of the light elements 
(helium, lithium, deuterium [an isotope of hydrogen] and others) in terms
of stellar nucleosynthesis and collision with cosmic rays in the remote past of the Universe.
The heaviest elements could also be explained in terms of stellar rather
than primordial nucleosynthesis, and the defenders of Big Bang in the end also had to adopt the
stellar nucleosynthesis of Burbidge et al.\ for the heavy elements.

Nonetheless, the steady state theory would lose competitiveness by the mid-sixties,
because it could not explain certain observational facts. It could not explain why the galaxies
were younger at higher redshift. It could not explain the excess of radio sources at large
distances (Ryle \& Clarke, 1961), nor the distribution of quasars. Most importantly,
it did not explain the cosmic microwave background radiation (CMBR), discovered in 1965
by Penzias and Wilson.\footnote{Indeed, the radiation had been discovered previously, but 
Penzias and Wilson, adviced by R. H. Dicke et al., interpreted it in cosmological terms
(Dicke et al., 1965). 
In the old Soviet Union, Shmaonov (1957) had made measurements at a frequency
of 9 GHz of a background radiation that was isotropic and had an antenna temperature of $4\pm 3$ K. There were also previous measurements by Japanese teams, and indirect measurements of the
existence of radiation of $\sim 2.3$ K by MacKellar in 1941 with the spectral analyses showing
excitation of rotational transition of cyan molecules (Novikov, 2001).} 
This strongly favoured the Big Bang theory.
 
In 1993--94, Hoyle, Burbidge, \& Narlikar (1993, 1994)\footnote{See also Hoyle, Burbidge, \& Narlikar (2000) or
 Narlikar et al. (2007) for a complete development of the theory and comparison with observational data.} 
 published a modification of the model that was called the ``quasi-steady state'' theory. The
main modification consisted in positing an oscillatory expansion apart from the exponential term:
\[
a(t)\propto e^{t/P}[1+\eta \cos(2\pi \theta(t)/Q)]. 
\]
$P\sim 10^{12}$ years, $\theta (t)\sim t$. 
The exponential factor had already been introduced in the first version of the Steady State model to keep
$\frac{\dot{a}}{a}=$constant and consequently maintain a constant density of matter by invoking the continuous creation of matter. 
The new term here is the sinusoidal oscillation.
The creation of matter is confined to epochs with minimum $a(t)$ rather than being continuous.
The parameter $Q$ and $\eta $ would be determined from Hubble's constant, the age of 
globular clusters and the maximum observed redshift in the galaxies. With this model,
some of the problems that affected the original theory of 1948 were solved.
This explained why there are younger galaxies at higher redshift, the problem of the 
radio sources, the distribution of quasars (with lower density for $z\gtrsim 2.5$),
the formation of large-scale structure (Nayeri et al., 1999).

The CMBR and its blackbody spectrum would be explained
as the effect of the thermalization of the radiation emitted by the stars of the last cycle
$P/3$ due to absorption and re-emision that produce needle-shaped particles (``whiskers'')
in the intergalactic medium. Due to the long distance travelled by the photons in the maxima
of the oscillation and due to the thermalization that occurs at each minimum, there is
no accumulation of anisotropies from one cycle to another. 
Only the fluctuations of the last minimum survive, which gives fluctuations of temperature
comparable to the observed $\Delta T/T \sim 5 \times 10^{-6}$. 
First, the carbon needles thermalize the visible light from the stars giving rise to 
 far infrared photons at $z\sim 5$, keeping the isotropy of the radiation. Afterwards,
iron needles dominated, degrading the infrared radiation to produce the
observed microwave radiation (Wickramasinghe, 2006). The anisotropies
of this radiation would be explained in terms of clusters of galaxies and other elements  (Narlikar et al., 2003, 2007).

Concerning the origin of the redshift in the galaxies, the proposers of this model admit a component
due to the expansion $a(t)$, like in the Big Bang, but furthermore they posit the existence
of intrinsic redshifts. This allows the solution of problems such as the periodicity of redshift in quasars, and the possible 
existence of cases with anomalous redshifts (L\'opez-Corredoira, 2010). 
The total redshift would be the product of both factors, expansion and intrinsic:
\[
(1+z)=(1+z_{\rm exp.})(1+z_{\rm int.})
\]

The intrinsic redshift is explained by means of the variable mass hypothesis. 
Hoyle \& Narlikar (1964) derived this hypothesis from a new gravitation theory based on Mach's principle with the solution that
the Minkowski metric and the particle mass depend on time as $m\propto t^2$. 
This variable mass hypothesis is used by the authors of QSSC to explain cases of 
anomalous redshifts, but it is not part of the main body of the hypothesis QSSC, that is, it is optional; QSSC can be conceived without the variable mass hypothesis.
The intrinsic redshift would be due to  variation of the energy of the emitted photon when
the masses of protons and electrons vary:
\[
(1+z_{\rm int.}) = \frac{m_{\rm observer}}{m_{\rm source}} = \frac{t_0^2}{(t_0-r/c)^2}
.\]

In the case of quasars, anomalies in the redshift would be observed because the mass
of their constituent particles grows proportionally to $(t-t_{quasar})^2$ instead of $t^2$ (Narlikar, 1977; Narlikar \& Arp, 1993).

Summing up, they proposed a model which aimed to compete with the standard ``Big Bang'' theory
but with a very different description of the Universe. According to the authors,
QSSC is able to explain the existing cosmological observations,
at least in an approximate way, and it can even explain some facts that the Big Bang model does
not explain (such as the anomalies in the redshifts of quasars). It also contains predictions
different from the standard model, though these are difficult to test. The predictions include
(Narlikar, 2006): existence of faint galaxies ($m>27$) with small blueshifts ($\Delta z<0.1$), 
the existence of stars and galaxies older than 14 gigayears, an abundance of baryonic matter
in ratios above those predicted by  the Big Bang, and gravitational radiation derived from the
creation of matter.

\subsection{Plasma Cosmology}

Plasma Cosmology assumes that most of the mass in the Universe is plasma controlled mainly by
electromagnetic forces (and also gravity, of course), rather than gravity alone, as in the standard model. 
The Universe has always existed, it is always evolving, and it will continue to 
exist forever. Some of its proposers are the Physics Nobel Prize laureate Hannes Alfv\'en (1908-1995), O. Klein, 
A. L. Peratt, E. Lerner, A. Brynjolfsson (Alfv\'en \& Klein, 1962; Alfv\'en, 1981, ch. 6; Alfv\'en, 1983, 1988; Lerner, 1991).

The plasma, through electric currents and magnetic fields, creates filaments similar to those
observed in the large-scale filamentary structure of the Universe. 
The  plasma cosmology model predicts the observer morphological hierarchy: distances
among stars, galaxies, cluster of galaxies, and filaments of huge sizes in the large-scale
structure. The observed velocities of the streams of galaxies in regions close to the largest
superclusters are coincident with those predicted by the model, without the need for dark matter (Lerner, 1991).
The formation of galaxies and their dynamics would also be governed by forces and interactions
of electromagnetic fields (Peratt, 1983, 1984; Lerner, 1991, chs. 1, 6).

Hubble expansion is admitted in the first version of plasma cosmology and was
explained by means of the repulsion between matter and antimatter.
Alfv\'en proposed his ``fireworks'' model in which a supercluster is
repelled by other superclusters; and within a supercluster each cluster is repelled by other
clusters; and within a given cluster each galaxy is repelled by the other galaxies, and so on,
obeying a distribution of matter and antimatter.
In each local volume, a small explosion would impose its own local Hubble relationship, and
this would explain the variations in the velocities of Hubble's law, i.e. the different values of the Hubble constant measured in the 70s and 80s, when Alfv\'en posited his hypothesis, in different ranges of distances or looking in different directions, all without invoking
dark matter. The energy derived from the annihilation of protons and electrons would produce
a background radiation of X- and $\gamma$-rays.
In more recent times, some proposers of plasma cosmology (e.g., Brynjolfsson, 2004; Lerner, 2006) 
have stated that there is no expansion, the Universe is static, and that the redshift of the galaxies would be 
explained by some kind of tired light effect of the interaction of photons with electrons in the plasma. 

With regard to the CMBR, Lerner (1988, 1995) 
explains it in terms of absorption and re-emission of the radiation produced by stars.
It is similar to the mechanism proposed by  QSSC, but here the
thermalization is due to interaction with electrons. The interaction of photons
and electrons produces a loss of direction in the path of the light, giving rise to
an isotropic radiation.

\subsection{Static Models}

There exist plenty of models which are characterized by lacking an origin of time (an eternal Universe), 
such as those described in the two previous subsections; such models moreover posit that
there is no expansion, in some cases the space even being infinite and Euclidean.
The redshift of the galaxies given by Hubble's law would be due to some mechanism different
from the expansion or Doppler effect, mainly a ``tired light'' effect (see reviews by L\'opez-Corredoira, 2003, \S 2.1; L\'opez-Corredoira, 2006). 
Among the many cases, I will mention just a few of them:

\begin{itemize}

\item The eternal Universe by Hawkins (1960, 1962a, 1962b, 1962c): Based on the
existence of a negative pressure in a cosmic fluid derived from general relativity
(not very different from the role the cosmological constant has acquired nowadays). 
The main point which differenciates this model from the standard theory is the proposal
that the Universe is static, infinite, without an instant of creation and without expansion.
The redshift of the galaxies is explained as a gravitational effect combined with a slight
amount of intergalactic extinction\footnote{Extinction produced by some particles which are placed in the space surrounding galaxies.} ($10^{-7}$ times the local interstellar absorption per unit
distance). Hawkins (1993) argues that his model is not unstable, with no tendency to collapse or expand, and that the CMBR is due to the emission of Galactic and intergalactic dust grains. 
Olbers' paradox (which says that integrating over infinite distance
we should get infinite flux) is solved by means of absorption in clouds of dust,
but energy does not disappear, so this dust should be heated and re-emit; this problem has no easy solution.

\item Chronometric cosmology (Segal, 1976; Segal \& Zhou, 1995): This model assumes that
global space structure is a 3D-hypersurface in a Universe of four dimensions.
Events in the Universe are ordered globally according to a temporal order.
The redshift of the galaxies obeys a quadratic law with distance (nowadays, it is
known this cannot be correct; Sandage \& Tammann, 1995). There is no explanation for the CMBR.

\item Curvature cosmology (Crawford, 2011; developed since the '80s and '90s): 
A new gravitational theory based on a combination of general relativity and quantum mechanics. 
The curvature pressure stems from the motion of charged particles in non-geodesic paths.
In the case of the photons that travel across the matter, this produces a ``tired light''
effect as a product of the gravitational interaction
between wave packets and curved space-time, giving rise to the observed redshift of 
galaxies. The result of the interaction of the photon is three new photons: one with almost 
identical energy and momentum to that of the original photon and two extremely low energy secondary photons.
Anomalous redshift cases might be produced by the extra redshift due to the photons'
passage through the cloud around the anomalous object (Crawford, 2011). The CMBR comes from the 
curvature-redshift process acting on the high-energy electrons and ions in the cosmic plasma. 
The energy loss which gives rise to the spectrum of photons of the CMBR occurs when an electron that 
has been excited by the passage through curved spacetime interacts with a photon
or charged particle and loses its excitation energy.

\item Wave system cosmology (Andrews, 1999): The Universe is a pure system of waves 
with mass density and tension parameter proportional to the local intensity 
of the modes of the waves. The peaks of the constructive interferences are the elementary
particles. The redshift is produced by a ``tired light'' mechanism. 

\item Subquantum kinetics (LaViolette, 2012) is a unified field theory with the foundations for a new wave theory of matter. Its non-dispersing, periodic structures resolve the wave-particle dualism and produce de Broglie wave diffraction effects.
Subquantum kinetics model proposes an open, order-generating universe, continuously creating matter and energy. It predicts that gravity potential should have a finite range. It uses ``tired-light'' redshift
in a static Universe, without radiating a secondary photon, no angular deflection, no strong wavelength dependence. It works as if intergalactic space on the average were endowed with a negative gravitational mass density.

\end{itemize}

\subsection{Variations on the Standard Model}

There are also models which are closer to the main characteristics of the standard model, but they are different in some minor aspects. 
Many of these models are investigated by some main 
stream cosmologists. They are alternative models which stem from the variations of the standard model. Here are some examples:

\begin{itemize} 
\item Newtonian cosmology: In the early development of Big Bang cosmology there appeared
a proposal (Milne, 1933, 1934) to keep an infinite euclidean space, with Newtonian gravity
and expansion as a pure Doppler effect in the  recession of the galaxies.
Many facts that were explained by the standard model with general relativity could also be explained with Newtonian cosmology. 
There remained some problems (stability,
Olbers' paradox), but there are also proposals to solve them without general relativity
(see the review by Baryshev \& Teerikorpi, 2012, \S 7.1.3)

\item The fractal Universe (e.g., Baryshev et al., 1994; Gabrielli et al., 2005): 
The density distribution of the Universe is not homogeneous on very large scales, but
obeys a fractal distribution. That is, the density within a sphere of radius $R$ is not
proportional to $R^3$ for large enough $R$ (in the regime in which there should be homogeneity) but
proportional to $R^D$ with a fractal dimension $D<3$.  

\item The cold Big Bang (Layzer, 1990; developed since the '60s): 
Rather than a very high temperature at the beginning of the Universe with a later progressive
cooling, the Universe starts with $T=0$ K. Alternative explanations are offered for
the origin of the elements (Aguirre, 1999), the CMBR (Aguirre, 2000) and
other phenomena explained by the standard hot Big Bang. 

\item Variations or oscillations of physical constants  ($c$, $G$, $h$, etc.) 
with time or distance.

\item Modifications of aspects of the gravity law. For instance, modified Newtonian 
dynamics [MOND, reviewed by Sanders \& McGaugh, 2002], which posits that Newton's law of gravitation 
is not followed for very low accelerations. Another such theory is modified gravity (MOG), and there are many other cases.

\item Multiple variations on the type of dark matter, dark energy/quint\-es\-sen\-ce, or even a 
Universe without these elements of the present-day standard model. For instance, some authors claim that
non-baryonic (cold) dark matter in haloes is not necessary to explain the rotation curves of the galaxies: 
with the above-mentioned MOND scenario, for example, dark matter is explained in terms of massive
photons (Bartlett \& Cumalat, 2011), protons and alpha particles moving at relativistic speeds
(so they interact very little; Drexler, 2005), magnetic fields (Battaner \& Florido, 2000), some
distribution of mass in the outer discs (Nicholson, 2003; Feng \& Gallo, 2011), etc.
Other examples are alternative proposals to explain the Hubble diagram of 
supernova data in terms other than the standard dark energy interpretation: an 
inhomogeneous Universe (Romano, 2007), evolution of SNIa luminosity (Dom\'\i nguez et al., 2000) 
or the absorption of their light by grey dust (Bogomazov \& Tutukov, 2011), intergalactic extinction, 
variation of $c$ and $G$ (as mentioned
in the previous point), other cosmologies, etc.

\item Multiple variations on inflation (alternative proposals such as cosmic strings, 
walls and other textures). Variations in the number of neutrino families; the formation of structures 
in a monolithic way (galaxies all formed at once) rather than the standard hierarchical scenario 
(the galaxies being formed in continuous episodes of accretion and merging), etc.

\end{itemize}

\subsection{Caveats/Problems in the Standard and Alternative Approaches}
\label{.caveats}

All models have gaps and caveats in their explanation of certain data derived from observations.
The Big Bang has a lot of problems and aspects that do not work properly or are not 
totally understood yet (see the reviews by L\'opez-Corredoira, 2003; 
Perivolaropoulos, 2008; Unzicker, 2010; Crawford, 2011; Kroupa, 2012; Baryshev \& Teerikorpi, 2012). 
Such problems include: higher metallicity and dust content at high redshift than expected,
much higher abundance of very massive galaxies at high redshift than expected, 
poorly understood extreme evolution of galaxy sizes, 
galaxies with $^4$He$<24$\%, ill-understood deuterium abundances, 
failure in the predictions of Li, Be, $^3$He, inhomogeneities at scales $>200$ Mpc,
periodicity of redshifts, correlations of objects with low redshift with objects at high 
redshift, flows of large-scale structure matter with excessive velocity,
an intergalactic medium temperature independent of redshift, a reionization epoch 
different from CMBR and QSO observations, anomalies in the CMBR (alignment 
quadrupole/octopole, insufficient lens effect in clusters, etc.), wrong predictions 
at galactic scales (no cusped halos, excessive angular momentum, insufficient number of satellites, etc.),
no dark matter found yet, excessive cluster densities, dark energy in excess of theoretical models by a factor $10^{120}$, 
no observation of antimatter or evidence for CP violation,
problems in understanding inflation, and so forth.

The expansion itself has no 
direct proof (nobody has directly observed a galaxy increasing its distance with respect to us); 
the most direct argument in favour of expansion is the redshift of the galaxies, but the redshift has possible 
explanations other than expansion. Most tests of expansion are dependent on the evolution of galaxies, so
they cannot give us a solution without a priori assumptions on that evolution.
There are a few tests which are dependent on other factors; for instance, the 
Alcock--Paczy\'nski test is independent of the evolution of the galaxies
but it presents entanglement on the cosmological effects with the 
redshift-space distortions (Ross et al., 2007).  
The CMBR, light element abundances and large scale structure formation
also have alternative explanations, as mentioned in previous subsections.
Other very recent fashionable topics in cosmology such as Baryonic Acoustic Oscillation (BAO) peaks
might be understandable in terms of different interpretations of the large scale structure too (L\'opez-Corredoira \& Gabrielli, 2013).
 
Of course, if the Big Bang model has problems, the alternative proposals have their own share of 
difficulties too, and their problems are more severe (see, for instance, Edward L. Wright's 
web-page\footnote{http://www.astro.ucla.edu/$\sim $wright/errors.html}), perhaps because these theories
are not as developed and polished as the standard model. For the expansion, either
they take it as fact, so they need speculative elements to argue that there was no 
beginning of the Universe (e.g., continuous creation of matter in QSSC) or an alternative explanation for the redshift of the galaxies. The CMBR has 
alternative explanations different from the Big Bang, 
but with some ad hoc elements (e.g., whiskers to thermalize 
stellar radiation in QSSC) without direct proof. Also, light element abundances require very 
old populations that have not been observed yet. 

Indeed, alternative models like QSSC do not apply 
a different methodology from the standard model:
both standard and QSSC models have some basic tenets and a lot of free parameters and ad hoc elements which are introduced every time some observation does not fit their models.
Its modern version (Hoyle et al., 2000) 
is able to explain most of the difficulties of the previous (steady state) version of the model. They introduce ad hoc elements without
observational support in the same way that the Big Bang introduces ad hoc 
non-baryonic dark matter, dark energy, inflation, etc. 
And they continue to skip the inconsistencies ad hoc: for instance, the maximum redshift of a galaxy was set to be 5 in the initial version of QSSC; however the authors have some free parameters which can be changed conveniently when some new observations do not fit the initial predictions, so at the end they can introduce ad hoc corrections which render their theory compatible with any maximum redshift of a galaxy. 
Indeed, something similar is done with the Big Bang theory: think, for instance, the predictions of the Big Bang for the maximum redshift of galaxy or the epoch of reionization.
They do the same kind of re-fitting of parameters.
Why, then, are the different theories accepted/rejected with different criteria?

The number of independent measurements relevant to
current cosmology and the number of free parameters 
of the theory are of the same order (Disney, 2007): in the '50s
the ``Big Bang'' was a theory with three or four free parameters to 
fit the few quantities of observational cosmology (basically, Hubble's constant 
and the helium abundance), and the increase in cosmological information from
observations, with the CMBR anisotropies and others, has been accompanied 
by an increase in free parameters and patches (dark matter, dark energy, inflation, initial conditions, etc.) 
in the models to fit those new numbers, until becoming today a theory with around 20 free parameters  
(apart from the initial conditions and other boundary conditions introduced in the simulations to reproduce 
certain structures of the Universe). A similar situation is given in particle physics too (Unzicker, 2010).

The number of independent measurements in CMBR anisotropies is also very limited.
While its power spectrum shows repeated information in the form of multiple peaks and oscillations, its Fourier transform, the angular correlation function, offers a more compact 
presentation that condenses all the information of the multiple peaks into a localized real space feature. 
Oscillations in the power spectrum arise when there is a discontinuity in a given derivative of the angular 
correlation function at a given angular distance (L\'opez-Corredoira \& Gabrielli, 2013). These kinds of 
discontinuities do not need to be abrupt over an infinitesimal range of angular distances but may also be smooth, 
and can be generated by simply distributing excesses of antenna temperature in filled disks of fixed or variable 
radii on the sky, provided that that there is a non-null minimum radius, and/or that the maximum radius is constrained. 
This allows a physical interpretation of these mathematical properties of CMBR anisotropies in terms of matter 
distribution in the fluid generating the radiation. A power spectrum with oscillations is a rather 
normal characteristic expected from any fluid with clouds of overdensities that emit/absorb radiation 
or interact gravitationally with the photons, and with a finite range of sizes and distances for those clouds (L\'opez-Corredoira, 2013a). 
The standard cosmological interpretation of ``acoustic'' peaks, from the hypothesis of
primeval adiabatic perturbations in an expanding universe (Peebles \& Yu, 1970), is just a particular 
case; peaks in the power spectrum might be generated in 
scenarios that have nothing to do with oscillations due to gravitational compression 
in a fluid. 

The CMBR angular correlation function can be fitted by a generic function 
with a total of $\approx$6 free parameters; saying that the power spectrum/angular correlation function 
contains hundreds or thousands of independent parameters for a given 
resolution is not correct, because their different values are not independent 
in the same sense that hundreds of observations of the position and velocity of a planet
do not indicate hundreds of independent parameters, the information of the 
orbit of planet being reduced to six Keplerian parameters. 
Nonetheless, the standard model with six free parameters (there are indeed $\sim$20 parameters, but the most 
important ones are six in number, the rest  producing low dependence) produces a still  better fit than 
the generic fit with the same number of free parameters; it fits third and higher order peaks whereas 
the generic fit reproduces only the first two peaks (L\'opez-Corredoira, 2013a).
There are also other theories that reproduce the same data with 
totally different cosmologies with a similar number of free parameters; e.g., Narlikar et al. (2003, 2007) 
for QSSC, Angus \& Diaferio (2011) for MOND. The fact that different cosmologies with different elements 
can fit the same data (with a similar number of free parameters to be fitted) indicates that the number 
of independent quantities in the information provided by the data is comparable to the number of free parameters in any of the theories.

There is near consensus in the values of the cosmological
parameters. The independent cosmological
numbers extracted from observations are of the same order.
Note, however, that there are some numbers which cannot be fitted. And
the publication of the measurements of these cosmological parameters may be biased 
due to the existence of values expected a priori. For instance, the analysis 
by Croft \& Dailey (2011) shows us that:
The value of the Hubble constant had a huge dispersion of values around two values of 
50 and 100 km/s/Mpc respectively before 1995, whereas immediately after 1995 almost all
 values clustered with small errors very close to the preferred value of 70 km/s/Mpc 
given by the HST Key Project; Before 1999, approximately 1/3 of the measurements of 
$\Omega _m$, using galaxy peculiar 
velocities, gave values inconsistent with being lower than 0.5 whereas after 1999, 
all measurements, including some using similar techniques, 
grouped around the preferred value of 0.25--0.30; The measurements of $\Omega_\Lambda $, 
which was considered null before the '90s, have now settled at 0.7 and since 1995 it presents a 
dispersion much lower than  expected statistically from the error bars, which
 means that either the error bars were overestimated, or that there is a bias in the publication 
of results towards the preferred value. Other examples could be given.

The development of modern Cosmology is somewhat similar to the 
development of the Ptolemaic epicyclic theory. However, in this race to build more 
and more epicycles, the Big Bang model is allowed to make ad hoc corrections and add more and
more free parameters to the theory to solve the problems
which it finds in its way, but the alternative models are rejected when the gaps
or inconsistencies arise and most cosmologists do not heed their ad hoc corrections.
Why are the different theories accepted/rejected with different criteria?

\section{The Difficulties in Creating Alternative Models: A Sociological/\-E\-pis\-te\-mo\-lo\-gi\-cal Model of How Modern Cosmology Works}
\label{.soc}

In my opinion, alternative models are not rejected because they are not potentially
competitive but because they 
have great difficulties in advancing in their research against the mainstream. 
A small number of scientists cannot compete with the huge mass of cosmologists 
dedicated to polishing and refining the standard theory.
The present-day methodology of research in cosmology does not favour the exploration
of new ideas. The standard theory in cosmology became dominant because 
it could explain more phenomena than the alternative ideas, but it is possible that 
partial successes have propitiated the compromise with a general 
view that is misguided and does not let other ideas advance that might be closer 
to a more correct description of the Universe. 

\subsection{Methodology of science}

Basically, there are two different methodologies to study Nature, both inherited from
different ways of thinking in ancient Greece: the rationalist--deductive method and
the empirical--inductive method (e.g., Markie, 2012).
\begin{description}

{\it 
\item[{\it The rationalist--deductive method:}]
This is the method devised by Py\-tha\-go\-ras and Plato. The pure relations of numbers in arithmetic and geometry are the immutable reality behind changing appearances in the world of the senses.
We cannot reach the truth through observation with the senses, but only through pure
reason, which may investigate the abstract mathematical forms that govern the world.
In this way of thinking, there is a predominance of creation of abstract theories, and mathematical modelling 
predominates over experimental and observational results.}
There are good cases of success using a rationalist--deductive approach.
An example within modern science Einstein's general relativity, which was
posited from aesthetic and/or rational principles in a time in which observational
data did not require a new gravity theory.
{\it In fact, observational tests proved this theory successful.
Present-day physics and cosmology are
partially Pythagorean when a theory is created before the observations.
It is also common among modern Pythagoreans to approve of statements such as the search for
beauty in a mathematical construction describing physical reality, or the Divine plan by
which the creator designed the Universe. The physicist--mathematician tries to achieve something close
to a mystical approach, tries to read into the Mind of God. Also, analogously with religion,
this extremely theoretical physics and cosmology can only be understood by a priestly elite 
able to think in four or more dimensions or in terms of similar abstractions. 
 
\item[{\it The empiricist--inductive method:}] As opposed to the preceding method, this one
points out that Nature should be known 
through observations and extrapolations of them. This is the Anaxagoras' method  
of how to know Nature. Aristotle uses both inductive and deductive methods, and he 
says that ``the mathematical method is not the method of the physicists, because
Nature, perhaps all, involves matter'' (Metaphysics, book II). 
Certainly, mathematics is useful for physicists, in spite of what was said by Aristotle, and this is clear 
since Galileo Galilei
put the bases of the scientific method, but I agree with the Greek philosopher that matter is not the same thing as mathematical entities.
Matter is not numbers, or geometry, or arithmetic, or the analysis of functions. Matter (or, better,  matter--
energy) is the component of the physical Universe, and this is what constitutes the reality of Nature to be studied by physical sciences.
The empiricism of Galileo Galilei might be an example within modern science, in the sense that observation and experimentation
are a requisite prior to theoretization, although all 
scientists, even Galileo, are also partly Pythagorean and all pythagoreans are in part empirists too.
These are extreme positions which cannot usually  be found in a pure form, but it is clear that, in some 
researchers, one of the trends dominates.}
But, apart from a few exceptions, the empirical--inductive method is more usual in science.
Dingle (1937) made an aggressive attack against the rationalist--deductive
method in favour of the empiricist--inductive method, with terms
such as ``paralysis of reason'', ``intoxication of the fancy'', 
`` `Universe' mania'', ``delusions'', ``traitors'', ``treachery''.
Robertson and de Sitter also favoured an empiricist
inductive science.
In my opinion, cosmology should be derived empirically by first taking the data without preconceived ideas, and then interpreting them from all 
possible theoretical viewpoints. Certainly, there are always prejudices and intuitions in our minds that push us towards certain avenues of research, but at least we should openly consider all the theoretical possibilities that can explain the
data, rather than taking only one (standard) theory and always trying to squeeze the data into it in some way.
In the words of Sherlock Holmes\footnote{The famous character 
of the novels by Arthur Conan Doyle.}:
`It is a capital mistake to theorize before you have all the 
evidence' (\textit{A Study in Scarlet}), and `before one has data, one begins to twist facts to 
suit theories instead of theories to suit data' (\textit{A Scandal in Bohemia}) [cited by
Burbidge, 2006].

\end{description}

Some astrophysicists closer to the observations than theory usually complain
about the lack of an empirical approach in cosmology. For instance, G\'erard de 
Vaucouleurs (1918--1995), known for his extragalactic surveys and Hubble's constant
measurements, said that there are  `parallelisms between modern 
cosmology and medieval scholasticism. (...) Above all 
I am concerned by an apparent loss of contact with empirical evidence and observational facts, 
and, worse, by a deliberate refusal on the part of some theorists to accept such results 
when they appear to be in conflict with some of the present oversimplified and therefore 
intellectually appealing theories of the universe' (de Vaucouleurs, 1970).
Certainly, the amount of data for observational cosmology nowadays is much
higher than in 1970 (although there were also many of them: CMBR, redshifts of galaxies, abundance of light elements, etc.); however, I think it is still valid nowadays: cosmology has not changed its methodology so much.

There is, however, an epistemological optimism encouraging the belief that successful theories
are successful because they reflect  reality in Nature. 
The philosopher of science Mosteir\'\i n (1989) said that scientists do not have any prejudice to 
accept alternative cosmologies. He also said, `there are no working alternatives to the standard
 big bang cosmological model (or family of models). This fact is not due to the will of the scientists 
 who created the model, still less to the prejudices of the scientific stablishment.
On the contrary, it is almost exclusively due to the strong observational constraints which 
reality puts on the activity of model-making. The standard big bang cosmological
model is the model no one wanted, but which recalcitrant experience forced everyone
to accept, at least for the moment being.' In my opinion, this kind of statement
is somewhat naive and denotes an excessive confidence in a fair application of scientific 
methodology. Certainly, all the available alternative models may be wrong, but this does not mean 
that they are rejected fairly; and  neither does it mean that the standard model is maintained for fair reasons.
This epistemological optimism might be correct in certain branches
of science but not in those areas close to metaphysical speculation such as cosmology, where
the scientific method is something like:

\begin{quotation}
--- Given a theory A self-called orthodox or standard, and a non-orthodox or 
non-standard theory B. If the observations achieve what was predicted by the 
theory A and not by the theory B, this implies a large success to the theory A, 
{\it something which must be divulged immediately to the all-important mass 
media. This means that there are no doubts that theory A is the right one. 
Theory B is wrong; one must forget this theory and, therefore, any further 
research directed to it must be blocked (putting obstacles in the way of 
publication, and giving no time for telescopes, etc.).}

--- If the observations achieve what was predicted by theory B rather than by 
theory A, this means nothing. Science is very complex and before taking a 
position we must think further about the matter and make further tests. It is 
probable that the observer of such had a failure at some point; further 
observations are needed 
{\it (and it will be difficult to make further observations 
because we are not going to allow the use of telescopes to re-test such a 
stupid theory as theory B). Who knows! Perhaps the observed thing is due to 
effect `So-and-so', of course; perhaps they have not corrected the data from 
this effect, about which we know nothing. Everything is so complex. We must 
be sure before we can say something about which theory is correct.} 
Furthermore, by adding some new aspects in the theory A surely it can 
also predict the observations, {\it and, since we have an army of theoreticians 
ready to put in patches and discover new effects, in less 
than three months we will have a new theory A (albeit with some changes) 
which will agree the data. 
In any case, while in troubled waters, and as long 
as we do not clarify the question, theory A remains. 
Perhaps, as was said by 
Halton Arp, the informal saying `to make extraordinary changes one requires 
extraordinary evidence' really means `to make personally disadvantageous 
changes no evidence is extraordinary enough'.} 
(L\'opez-Corredoira, 2008)
\end{quotation}

Halton C. Arp (1927-- ), a heterodox observational cosmologist, known through
his proposal of non-cosmological redshifts (L\'opez-Corredoira, 2003, \S 2.8),
would point out: `Of course, if one ignores contradictory observations, 
one can claim to have an ``elegant'' or ``robust'' theory. But it isn't 
science.' (Arp \& Block, 1991)

\subsection{The Snowball Effect}

The alternative models try to compete with the standard model, but cumulative
inertia gives a clear social advantage to the standard model. This advantage determines
that researchers may continue to explore these alternative ideas. 
Metaphorically, it is like a snowball effect:
`The snowball effect arising from the social dynamics of research funding drove more 
researchers into the Standard Cosmology fold and contributed to the drying out 
of alternative ideas' (Narlikar \& Padmanabhan, 2001). 
It is not strange that Jayant V. Narlikar (1938-- ), 
one of the creators of the QSSC who still tries to keep it alive,
should be frustrated in his odyssey and should link the lack of social success of his theory to how social dynamics works. Anyway, regardless of his frustration, either from dissidence or orthodoxy,
what he claims is basically correct and applicable to most speculative sciences or half-sciences
such as cosmology. Another creator of the quasi-steady state, Geoffrey R. Burbidge (1925--2010),
did not have a higher opinion:

\begin{quotation}
Let me start on a somewhat pessimistic note. We all know that new ideas and revolutions 
in science in general come from the younger generation, who look critically at the contemporary 
schemes, and having absorbed the new evidence, overthrow the old views. This, in general, is the 
way that science advances. However, in modern astronomy and cosmology, at present, this is 
emphatically not the case. Over the last decade or more, the vast majority of the younger 
astronomers have been conformists in the extreme, passionately believing what their leaders 
have told them, particularly in cosmology. In the modern era the reasons for this are even 
stronger than they were in the past. To obtain an academic position, to obtain tenure, to be 
successful in obtaining research funds, and to obtain observing time on major telescopes, it 
is necessary to conform. (G. R. Burbidge, 1997)
\end{quotation}

Here is a similar opinion from a researcher who is not particularly heterodox: 

\begin{quotation}
It is common practice among young astrophysicists these days to invest
research time conservatively in mainstream ideas that have already been explored
extensively in the literature. This tendency is driven by peer pressure and job market
prospects, and is occasionally encouraged by senior researchers. Although the same
phenomenon existed in past decades, it is alarmingly more prevalent today because a
growing fraction of observational and theoretical projects are pursued in large groups
with rigid research agendas. In addition, the emergence of a 'standard model' in
cosmology (albeit with unknown dark components) offers secure 'bonds' for a safe
investment of research time. (Loeb, 2010)
\end{quotation}

The snowball effect, also called Matthew effect (Merton, 1968)\footnote{Merton (1968) gave it the name ``Matthew effect'' from the Gospel of St. Matthew (25:29):
`Unto every one that hath shall be given, and he shall have abundance:
but from him that hath not shall be taken away even that which he hath.'},
is to a certain extent present in the social dynamics of cosmology, as well as in 
other speculative areas of science (L\'opez Corredoira, 2013b, \S 3.8).
It is a feedback ball: the more successful the standard theory is, the more money and
scientists are dedicated to work on it, and therefore the higher the number of observations
that can be explained with more parameters and ingredients (dark matter, dark energy, inflation, etc.) 
introduced ad hoc, and that cause the theory to be considered more successful.

However, not everything is a social
construct (as some postmodernists claim): the CMBR, the redshift of
galaxies, etc. may be real facts, or at least I have no doubts of their existence although other
authors have expressed such\footnote{For instance, there are some authors (Li et al., 2009; Cover, 2009) who suspect that all of the reduction of raw data of CMBR have common a priori assumptions which lead to the same measurement of power spectrum, but it could change or even be compatible with no anisotropies with different methodology applied to the analysis of raw data.}, 
and they also have weight in the credibility of the standard model.

\subsection{Censorship and {\it arXiv.org}}

It is also worth noting that the publication of heterodox ideas is far to be free, in particular
in recent years. Apart from the refereed journals, which usually reject challenging ideas deviating
from mainstream points of view, there is another important tool for communicating
scientific results in physics: the preprint server {\it arXiv.org}. It is a monopoly within physics and 
has no competitors. Even most of the papers published in journals are posted on this preprint
server, and people read them here. 
The situation is that papers not posted on {\it arXiv.org}, 
will receive scant dissemination within the community, particularly when the papers are
not published in a reputed refereed journal, which is often the case for non-mainstream positions.

The development of {\it arXiv.org}, first at Los Alamos National Laboratory and later at Cornell University, 
was a wonderful example of freedom of expression between 1992 to 2004 that provided everybody with an open
forum in which to post their ideas. There was a small fraction of papers with 'exotic' ideas, but they were
very few (5\% or less), so they did not disturb the flow of information. 
However, after 2004 there was a change in policy and those responsible for the site decided to
block the posting of certain contributions. In 2004, a system was introduced  in which in order to post something on the site support was requested from a colleague with experience in the field. 
The methods of the system would become more subtle in the following years, forbidding  some scientists from giving support when arXiv moderators noted that they had allowed the publication of very challenging heterodox ideas, and creating committees to reject  papers 
without having read them and with the absence of a referee's report: the committees just read the title and the abstract and, if they did not 
like the content (and normally they do not like anything that has not been accepted in a refereed journal and smells of heterodoxy, such as denial of the expansion of the Universe or discussions about alternative interpretations of the CMBR), they channel the paper, which formerly would have been placed on 'astro-ph.CO', widely read by astrophysicists, to 'physics.gen-ph', which is hardly read by anybody. In some cases, they remove the contribution totally, without further explanation (e.g., Castro Perelman, 2008).
When asked for an explanation for a rejection, they usually reply with set phrases: 
`arXiv reserves the right to reclassify or reject any submission. We are not obligated to provide substantive 
reasons for every rejection, and usually the moderators do not provide more than a sentence or two, often in 
a form not appropriate for author viewing'.
This method of censorship of the promotion of new ideas 
in cosmology appears to me to be somewhat on a par with certain totalitarian regimes
(see further discussion in L\'opez Corredoira, 2013b).

{\it 
\subsection{The Influence of Culture and Religion}

Another factor that carries some weight in the determination of the dominant scenario in cosmology
is the ideology of the researchers, and in the case of religious ideas this is somewhat relevant.
The association of cosmology and religion is indeed very old--says Kragh (2007b)--but there
are in my opinion older themes that are never overcome.

In \textit{Timaeus}, Plato says that time was created simultaneously with the Universe. 
This idea was introduced into Christianity from
the third century A.D., after reconciling Christianity with existing Roman society
and its ideas influenced by Plato and Emperor Tertullian (Roberts, 1924; Lerner, 1991, ch. 2).
Augustine of Hippo later introduced certain Platonic
ideas into Christianity, such as the untrustworthiness of the senses and the instantaneous
creation of the Universe from nothing.
A universe of infinite space and time is exclusive to the Deity,
and thus prohibited for the material universe.

The astrophysicist Binggeli (2006) compares the standard model of modern cosmology with the 
cosmology in the Judaeo--Christiano--Gnostic beliefs of the Scholastic Middle Age, depicted in 
Dante's Divine Comedy (\textit{Primum Mobile}), and the author finds that there is a perfect correspondence 
in some essential points between both worldviews. The three basic tenets of \textit{Primum Mobile} are 
present in the observable Universe of the Big Bang theory: 1) there is a maximum finite distance 
from us in the observable Universe, 2) the observable Universe is a sphere with us at the centre; and 3) 
it has a hierarchical structure. One may wonder about the cause of these correspondences, and 
the answer is also given by Binggeli (2006): there must be a psychological 
mechanism dominating our visions of Nature. The result of our research is not objective but 
highly biased by the influence of the culture in which we are embedded (which has inherited the 
Scholastic cosmological view) and our own psychological patterns. Modern cosmology is a symbolic 
expression of the states of our mind. The author argues that our view of the external reality is 
indeed a reflection of our interior world, and that the way to understand modern science should 
go through a psychological analysis. I think that in some degree he is right: cosmology depends 
on the social and psychological conditions of scientists. Nonetheless, we should not forget that 
there are also some elements that are not a reflection of our souls but that result from the observation of something which is outside us.

Because of this historical background and the coincidences of elements of
the standard model with certain credos, some authors think that nowadays the Big Bang is simply the
scientific version of Genesis, and that to many people, the Big  Bang idea is attractive
in the same way, being a synthesis of astrophysics and the dogma of a
creation ex nihilo (e.g., Jastrow, 1978). Indeed,
in 1951 (when the Big Bang was not yet a dominant standard theory), 
Pope Pius XII asserted that the Big Bang
supports the doctrine of creation ``ex nihilo'' (Pius XII, 1952). 
He wrote in an address to the Pontifical Academy of Sciences:

\begin{quotation}
In fact, it seems that present-day science, with one sweeping
step back across millions of centuries, has succeeded in bearing
witness to that primordial `Fiat Lux' (Let there be light) uttered at the
moment when, along with matter, there burst forth from nothing a sea
of light and radiation, while the particles of the chemical elements
split and formed into millions of galaxies... Hence, creation
took place in time, therefore, there is a creator, therefore, God exists!
\end{quotation}

In 1982, a conference on cosmology was held at the Vatican. The conference
was confined completely to Big Bang cosmology and its proponents; radicals
such as F. Hoyle, V. Ambartsumian and G. Burbidge were not invited.
Many prestigious scientists have also used the ideas of modern cosmology for
theological claims. There are many who talk about the Big Bang leading to 
a proof of God's existence (e.g., Davies, 1983; the debate for and against the idea 
in Soler Gil \& L\'opez Corredoira, 2008). George F. Smoot, when the discovery of
the anisotropies of the CMBR 
were announced, claimed that for a religious
person this was looking at the face of God (Wright, 1993). 
We must also bear in mind that the United States, at present the leading country in cosmological research, 
is dominated by a much higher proportion of followers of the Christian religion than in
other rich countries.

Christianity is not the only religion to have found this association of concepts.
There also seems to be
great acceptance of the standard cosmology in other monotheistic religions. The Israeli
physicist and cosmologist Moshe Carmeli (2000) says that not only does the Big Bang
scenario agree with the idea of creation described in the Bible, but also with the
scenario of creation in six days. The Muslim astronomer Kamel Ben Salem (2005) analyses 
the Quranic description of phenomena linked to the evolution of the universe.

The opposite trend is also observed. Among heterodox scientists and sceptics (myself included) 
there has been and continues to be a higher ratio of atheists and agnostics. It is known, for instance, that 
Fred Hoyle was not a believer. And there are cases of practices in communist countries that
favoured non-standard cosmologies.
For instance, in the People's Republic of China till the '70s, there was some degree of censorship affecting
the circulation of ideas relating to the Big Bang (Hu, 2004; Kragh, 2007a, pp. 199-200).

Of course, there are also many atheists who follow Big Bang and vice versa, but there have 
been correlations between religious dogma and preferred cosmological scenario and these correlations
are not fortuitous. This makes us appreciate the weight of ideology in the early development of
scientific ideas such as cosmology, i.e. that cosmology is not totally objective. Nonetheless, from
what I can observe among my colleagues (almost all cosmologists, either christians or atheists, are pro-Big Bang) once the standard paradigm is in a dominant position, religious ideas do not exert such a strong 
influence, and other sociological factors seem to be more important.

}

\subsection{The Psychological Profile of Cosmologists}

Social trends or ideologies can greatly influence the kind of science that is carried out in a given
epoch and the corresponding results.
Also, at the level of the individual, the psychological profile
of the researcher can produce leanings towards either orthodoxy or heterodoxy.
In my experience, cosmologists tend to fall in one of the following extreme categories, with gradations of grey between them:

\begin{description}

\item[Heterodox:] possessed by the complex of unappreciated genius, 
{\it too much ``ego'' which does not discourage the researcher in the difficulties for the
creation of a new alternative model.}
Normally working alone/in\-di\-vi\-dual\-ly or in very small groups,
creative, intelligent, non-conformist. His\footnote{As far as I know, there are no women
doing this kind of research with their own global cosmological model. 
If somebody knows any exception, let me know it.}
dream is to create a new paradigm in science which completely changes our view 
of the Universe. Many of them try to demonstrate that Einstein was wrong, 
maybe because he is the symbol of genius and defeating his theory would mean that
they are geniuses above Einstein. But they are not what they think they are, and
most of their ideas are ill-founded.
{\it Most of them are crackpots with crazy ideas with
little to be said in their support.}
Few of them need to be taken seriously.

\item[Orthodox:] dominated by groupthink,\footnote{In a sociological analysis, 
Janis (1972) categorizes the symptoms of groupthink as: 1) An illusion of invulnerability, 
shared by most or all the members, which creates excessive optimism and encourages the taking of 
extreme risks. 2) An unquestioned belief in the group's inherent morality, allowing 
the members to ignore the ethical or moral consequences of their decisions. 3) Collective efforts 
at rationalization in order to discount warnings or other information that might lead the members 
to reconsider their assumptions before they recommit themselves to their past policy decisions. 4) 
Stereotyped views of enemy leaders as too deviant to warrant genuine attempts to negotiate, or as 
too weak and stupid to
counter risky attempts made at defeating their purposes. 5) Self-censorship of deviations from the
apparent group consensus, reflecting each member's inclination to minimize to himself 
the importance of his doubts and counterarguments. 6) A shared illusion of unanimity concerning 
judgments conforming to the majority view (partly resulting from self-censorship of deviations, 
augmented by the false assumption that silence means consent). 7) Direct pressure on any member 
who expresses strong arguments against any of the group's stereotypes, illusions, or 
commitments, making clear that this type of dissent is contrary to what is expected of all loyal 
members. 8) The emergence of self-appointed mindguards - members who protect the group 
from adverse information that might shatter their shared complacency about the effectiveness and 
morality of their decisions (Dolsenhe, 2011, ch. 12). Sanrom\`a (2007) applied the concept of groupthink 
to present-day cosmology.} following a leader's opinion,
{\it as in the tale of the naked king. 
Any crazy opinion can be accepted if it is supported by the leading cosmologist, and in this
sense Big Bang theory, even if it is a very speculative set of hypotheses, still finds a place in
the psychology of the wider community of scientists and grow by the snowball effect.}
They are good workers, {\it conformist, domestic, 
performing monotonous tasks without ideas}
in large groups, specialists in a small field which they know 
very well, 
and in which they do not try to develop new paradigms. 
{\it His/her dream is getting a permanent position 
at an university or research centre, to} dedicate large portions of their time to the administration and 
politics of science {\it (i.e. astropolitics;} see L\'opez-Corredoira, 2008; 2013b, chs. 3, 6), {\it to be leader of a project. Many of them are like sheep (or geese\footnote{{\it In 
Hoyle et al. (2000), a serious and technical book about cosmology,
a picture was inserted in which a row of geese are turning around a corner all 
in the same way, with the following ironic comment:
``This is our view of the conformist approach to the standard
(hot big bang) cosmology. We have resisted the temptation to name some of the leading
geese''.}}), some of them have the vocation of shepherds too.}

\end{description}

The sociological reasons for favouring  orthodox proposals might be related to the
preference of domesticity in our civilization (see L\'opez-Corredoira, 2013b, ch. 5). 
An anarchy in which everybody expresses his or her ideas freely
is not useful for the system. 
{\it Sheep} rather than crackpots are preferred.
Finding a promising change of paradigm closer to the truth among thousands of crazy 
proposals is very difficult. In orthodoxy, although  absolute truth is not guaranteed, 
at least a consensus version of the truth is offered and that is what has weight.
By means of it, if somebody is wrong then everybody is wrong and the fault is diluted among
 many. Investment in science prefers security, it prefers domesticity and
control, rather than a promising and challenging change of paradigm that is uncertain, with the attendant
difficulty of guessing from which direction a new paradigm could come.
Nonetheless, again I insist, we must not forget that there is empirical evidence in favour
of the standard theory. Nature is more than a social construct or similar kinds
of postmodern claims.

\subsection{An Illustrative Example for the Sociology of Cosmology}

Somebody may think that the arguments given in this paper are just pure abstractions.
They are not, they are based on observations of real cases. Perhaps
the case of a recent experience of mine might be illustrative. 

Every year in my research centre (the 'Instituto de Astrof\'\i sica de Canarias' {\it [IAC]}), there is a call
for proposals for the following year's Winter School for doctorate students and young postdocs. 
I have submitted a proposal with the title ``Different Approaches to Cosmology''
{\it and the following abstract: 

\begin{quote}
The aim of this winter school is to present the status of current 
cosmology from both a standard and non-standard points of view, discussing successes and failures. 
In particular, the standard model and a number of non-standard models will be presented
to provide the students with a set of tools to carry out and/or devise new experiments to
challenge the current paradigm, either to prove or disprove it. Particular emphasis will
be given to the comparison between prediction of the different models and observations.
\end{quote} }
I also included a list of possible speakers.
The topic attracted attention, so I was advised of the interest for the school, which
was chosen as the first option among the proposals, provided that the following changes were made:
{\it The first thing that I was told was that I should include the name of some women among
the list of possible invited speakers for political correctness of gender balance.
I replied that there are no women with their own alternative
theories in cosmology, but that we could include some  to talk about variations on the Big
Bang, or partial aspects of an alternative theory. 
The second complaint was that there should be a higher ratio of orthodox cosmologists
in the school, at least three or four of the total of eight speakers. I accepted this suggestion too.}
The last stage came when I sent my list of speakers with the eight names and possible replacements as second options (in total there were sixteen names) including both women and many orthodox cosmologists. 
{\it Many of the names were accepted but I received the following new complaint from the head of the 
research division at the IAC (the original e-mail on 31 October 2011 was in Spanish):}

\begin{quotation}
I have looked into it further, and I had the luck to get the comments
of a very senior astronomer who does not work directly in cosmology (which
is an advantage, because he/she is not set in his/her own scientific ways) and
has got a very wide experience in editing journals and organizing congresses. (...)
[My contact] told me that we could invite (...), but not 
{\it Yurij [Baryshev]}, 
not even
as a second option. (...) I like
{\it Eduardo [Battaner]} 
very much, but he is not
the appropriate person for the topic of magnetism in cosmology. (...) We cannot invite
{\it Arp}, 
as he is confrontational (...) [his recent
work] lacks a scientific basis. At this point I vetoed it. (...) This topic
[Plasma Cosmology and a proposal to invite 
{\it Eric J. Lerner}] 
is too marginal, and
I propose to forget it. (...) 
[{\it Jayant Narlikar} and the QSSC] 
No, I vetoed that too. He is totally marginal and the theory is dead.
CDM has its problems, but QSSC is not going to solve them. The only thing
we could do would be to invite 
{\it Simon White }
to tell us why the theory does not work,
or even organize a mini-debate between perhaps 
{\it Kroupa}
and another researcher about the topic.
\end{quotation}

{\it 
A magnificent example of how cosmology works. 
A school describing 
the most important ideas about cosmology, Big Bang and others, was proposed
and the idea had been initially accepted as interesting, as a sign of openness of the mind of our
scientific community. But what happened? When one gives some names of some of the most
important creators of heterodox ideas (Baryshev, Battaner, Arp, Lerner, Narlikar)
they were rejected because some members of a committee who were not even cosmologists
decreed that these theories were marginal and dead. 
The name of Virginia Trimble was also rejected for different reasons. }

The theories may be marginal and dead but not because irrefutable
scientific arguments against them were given, but rather precisely because  of this kind of attitude
in the organization of social scientific events (journals, meetings, etc.). 
Alternative theories die because they are being killed by the
same people who say that they are dead. 
{\it And most of the scientists who claim that these theories 
are dead/marginal have never read a paper on these ideas and they merely repeat what they have 
heard from some colleague (groupthink, blindly following the opinion of the leaders).}
What was particularly shocking was the
rejection of the invitation to 
{\it Narlikar} 
on the grounds that the QSSC theory is dead.
{\it Indeed,
what the censors probably meant is that Fred Hoyle (1915--2001, father of the idea) 
and Geoff Burbidge (1925--2010, a physicist with an important influence in the political decisions on
astrophysics; former editor of the highest impact factor journal of astrophysics, the \textit{Annual
Review of Astronomy and Astrophysics}) are dead, certainly, so it is understood
that there is now no living sacred cow to respect, and the community decides
to declare that the theory is dead.} 
However, I have not seen any scientific paper in the last
decade that demonstrates irrefutably that the basic points of the QSSC are untenable.
The final suggestion was the most revealing one: that we invite 
{\it Simon White} 
to tell us why QSSC is wrong, without giving 
{\it Narlikar}
the chance to defend his ideas.
This is equivalent to organizing a meeting on different religious ideas
and inviting only christians to participate, including a speech of the 
Pope to tell us why hinduism is a false doctrine. 
Finally, my proposal was rejected because I insisted in keeping at least
two of the five rejected names.

\section{Limits of Cosmology}
\label{.lim}

\begin{quotation}
And we would pretend to understand everything about cosmology, which concerns the whole 
Universe? We are not even ready to start to do that. All that we can do is to enter in the field 
of speculations. So far as I am concerned, I would not comment myself on any cosmological theory, 
on the so-called `standard theory' less on many others. Actually, I would like to leave the door 
wide open.  [Jean-Claude Pecker, in Narlikar et al. (1997)]
\end{quotation}

I agree with Jean-Claude Pecker (1923-- ), 
another classical heterodox dissident cosmologist.
Before wondering which is the true model of cosmology, we must wonder whether we are
in a condition to create a theory on the genesis (or non-genesis) and evolution of the
whole Universe, whether the psychologico--so\-cio\-lo\-gi\-cal conditions of the cosmologists
are or are not weightier factors than observations of Nature.
Is present-day cosmology dominated by our culture or by Nature's objective truths?

\subsection{The dogma of the cosmologist, according to Mike J. Disney}

According to Michael J. Disney (1937- ), in his brave paper 
`The case against Cosmology' (Disney, 2000), present-day cosmology is a dogma with
a serie of gratuitous or quasi-gratuitous assumptions: 

\begin{description}

\item[The non-theological assumption:] speculations are not made which cannot,
at least in principle, be compared with observational or experimental data,
for tests.

\item[The ``good-luck'' assumption:] the portion of the Universe susceptible to
observation is representative of the cosmos as a whole.
 
\item[The ``simplicity'' assumption:] the Universe was constructed using a
significantly lower number of free parameters than the number of clean and
independent observations we can make of it.

\item[The ``non-circularity'' assumption:] the Laws of Physics which have significantly
controlled the Universe since the beginning are, or can be, known to us from
considerations outside cosmology itself.

\item[The ``fortunate epoch'' assumption:] we live in the first human epoch
which possesses the technical means to tease out the crucial observations. 
This is also expressed by Narlikar (2001): `there is one trait which
the cosmologists of old seem to share with their modern counterparts,
viz. their fond wish that the mystery of the nature of the universe would
be solved in their lifetime.'

\end{description}

From this, Disney (2000) concludes:

\begin{quotation}
I can see very little evidence to support any of the last 4 assumptions while it is dismaying to find that some cosmologists, who would like to think of themselves as scientific, are quite willing to abrogate the first.
\end{quotation}

He also says in another part of his text, referring to the cosmologists who think that
they can establish a cosmological model as securely as the Standard Model of elementary particles:
\begin{quotation}
We believe the most charitable thing that can be said of such
statements is that they are naive in the extreme and betray a complete
lack of understanding of history, of the huge difference between an
observational and an experimental science, and of the peculiar limitations
of cosmology as a scientific discipline.
\end{quotation}

That is the extremely sceptical position of an astrophysicist with a long career who has made significant
contributions to extragalactic astrophysics. We may interpret it as too daring, as an
exagerated parody that is out of place in the present cosmological scene. \'Cirkovi\'c (2002) criticizes Disney (2000) saying that his claims are rhetorical with no new ideas about the sociology/philosophy 
of science, and that his critique is unfair, biased and constrained in an extreme inductivism.
Other disciplines operate in a similar way to cosmology and they are sciences, says \'Cirkovi\'c.
But we could also pay attention to some of Disney's sentences and see that there is some 
background of truth in what he claims, in spite of the exaggeration.

\subsection{Is a Science of Cosmology Possible?}

I would say that
before understanding the Universe, we must understand the pieces of the puzzle separately
(galaxies, their formation, their evolution, whether they separate from each other, the origin
of the elements, the origin of the CMBR) rather than assembling all of
them into a happy idea that could convert astrophysics into a speculative science.
There are however many cosmologists, philosophers and historians of science who think
that cosmology became an empirical science beyond speculation after the discovery of the CMBR (e.g., 
Kragh, 2007b). As I have maintained throughout this paper, 
I do not agree with Kragh's (2007b) statement that cosmology is a proper science like nuclear physics,
hydrodynamics, etc. Even if there are aspects which are comparable with observations, they are just a
few partial aspects of the whole reality, whereas cosmology stands for a science of the
whole Universe and its whole history, something for which we do not have all the empirical/observational 
information that we need to have to fill in the many gaps in that history that are so far questions of pure
speculation and risky extrapolations. 

Is cosmology comparable perhaps to palaeontology or a science which
tries to reconstruct the facts from fossils (\'Cirkovi\'c, 2002)? No, I do not think it can
be put on the same level of scientificity as palaeontology, because the objects of study in
palaeontology are much more limited, and the geological and biological processes are known, whereas
cosmologists play with elements for which there is no direct experience (dark..., 
dark..., new physics...), or they must adopt extrapolations and assumptions for which there is no
evidence (the cosmological principle, the principle that the laws of physics do not change over time, etc.).
This means that the process of choosing 
between standard and non-standard models in cosmology is less fair
(less based on evidence) than in other scientific disciplines.
In any case, there are also huge extrapolations involved in disciplines such as 
palaeogeography and palaeobiology. Certainly, one can doubt the different theories of different fields of science, but for different reasons.
The very word ``Universe'' also merits some consideration: it means everything 
that exists or has existed, and we have access only to a small part of the observable cosmos
(the ``good luck'' assumption given in the cosmologist's dogma by Disney, 2000).
In this aspect, palaeontology is not so different from cosmology because it  only
has access to a small part of the observable universe.

As Kragh (2007b) remarks (as an argument
considering cosmology as a science), Nobel prizes have been given to some cosmologists. In my opinion,
this does not mean anything. Nobel prizes are just part of sociological structures. Recognition
does not mean a higher value of some knowledge and its creators, but only higher status.
{\it Indeed, there are many social and economic interests in declaring cosmology to be a solid science,
there is a lot of money in the game, and this motivates the arrogance of the claim that we can
know the whole Universe and its history.}

\section{Conclusions}

Alternative theories are not at present as competitive as the standard model in cosmology.
If they were more developed, there is a possibility that they might compete in some aspects with
the Big Bang theory, but efforts are made in the present-day scientific community to avoid their
development. The fact that most cosmologists do not pay them any 
attention and only dedicate their research time to the standard model is to a great extent due to 
a sociological\footnote{For further reading on my impressions about sociological aspects of science in general, see L\'opez Corredoira (2013b).} 
phenomenon (the ``snowball effect'' or ``groupthink''). Cosmology, knowledge of the 
Universe as a whole, shares some characteristics with other sciences, and there is some scientific 
content in it. However, in my assessment, cosmology is more affected than most other sciences by human factors (psychological, sociological, ideologies/culture, etc.).

Note that I am not defending any specific dogma here: neither the correctness nor the 
wrongness of Big Bang; neither am I defending constructivism or scientific realism (see 
discussion on these positions, for instance, in Soler Gil, 2012). I am just presenting some sceptical 
arguments expressing certain doubts on the validity of the standard cosmology, 
and this requires seeing the problem from several points of view:
in this paper I have talked more about the social aspects and the alternative models.
Nonetheless, there are also reasons to support the standard model in a realistic way.

There are limits to cosmology because we are finite human beings limited by our experiences
and circumstances, not mini-gods able to read the mind of a god who played maths with
the Universe, as some Pythagoreans may think. There is a lack of humility in Pythagoreanism,
or in expressions like ``precision Cosmology''. One of the most reputed 
physicists of the former Soviet Union,
Lev Lavidovich Landau (1908--1968),
said: `Cosmologists are often in error, but never in doubt.'
Great old masters, even the creators of the standard model, were cautious
in their assertions. Edwin P. Hubble (1889--1953) throughout his life doubted the reality 
of the expansion of the Universe. Willem de Sitter (1872-1934) claimed: 
`It should not be forgotten that all this talk about the universe
involves a tremendous extrapolation, which is a very dangerous operation' (de Sitter, 1931).
This scepticism is sane since `all cautions are too little' (Spanish proverb).
It is not a question of substituting one model for another, since it would be the `same
dog with a different collar' (another Spanish proverb) but of realizing the limits
of cosmology as a science. 

Rutherford (1871--1937) said `Don't let me hear anyone use the word `Universe' 
in my department.' In the same style, the astrophysicist 
Mike Disney (1937-- ) 
dared to claim: `The word `cosmologist' should be expunged from 
the scientific dictionary and returned to the priesthood where it properly belongs' 
(Disney, 2000). Those are the words of an old-style scepticism. Nowadays, the young bloods of precision
cosmology do not care for such statements and  proudly claim  that people in the past did not know what they know.
Cosmologists with no indication of doubt and an amazing sense of security who
dissert on topics of high speculation. Of course, science advances, and 
cosmology advances in the amount of data and epicycle-like 
patches to the theory to make it fit the data, but the great questions remain 
almost unchanged. Many wise men have already deliberated on cosmology for a
long time, without reaching a definitive solution.
Do we live in a fortunate golden age of cosmology that allows us, 
thanks to our technical advances and our trained researchers, 
to answer questions on eternity, the finiteness of the Universe, etc.? 
We could reply as the 19th century German philosopher Schopenhauer
did with the Know-alls of his time:

\begin{quotation}
Every 30 years, a new generation of talkative candid persons, ignorant
of everything, want to devour summarily and hastily the results of human
knowledge accumulated over centuries, and immediately they think themselves
more skilful than the whole past.
\end{quotation}
 
\

{\bf Acknowledgements:} Thanks are given to the editor of this special issue
on ``Philosophy of Cosmology'', Henrik Zinkernagel, and to the two anonymous 
referees of this paper for their helpful comments.
Thanks are given to Terry J. Mahoney (IAC, Tenerife, Spain)
for proof-reading and comments on this paper. The author was supported
by the grants AYA2007-67625-CO2-01 and AYA2012-33211 
of the Spanish Science Ministry.

\end{document}